%% file: paper_su2_rho.tex
 \documentclass[11pt,a4paper]{article}
    \usepackage{jheppub}

\usepackage{longtable}
\usepackage{amsthm}
\usepackage{amssymb}
\usepackage{amsfonts}
\usepackage{bbm}
\usepackage{color}
\usepackage{dcolumn}% Align table columns on decimal point
\usepackage{bm}% bold math
\usepackage{pxfonts}
\usepackage{slashed}
\usepackage{graphicx}
\usepackage{multirow}
\def\wick#1{\setbox2=\hbox{$\displaystyle#1$}
    \setbox3=\null\ht3=3.0pt\dp3=0.0pt\wd3=20.0pt
    #1\kern-\wd2\kern3.0pt\raise11.0pt\vbox{\hrule height0.3pt
    \hbox{\vrule width0.3pt\box3\vrule width0.3pt}}\kern-24.0pt\kern\wd2}

\def\longwick#1{\setbox2=\hbox{$\displaystyle#1$}
    \setbox3=\null\ht3=3.0pt\dp3=0.0pt\wd3=27.0pt
    #1\kern-\wd2\kern3.0pt\raise11.0pt\vbox{\hrule height0.3pt
    \hbox{\vrule width0.3pt\box3\vrule width0.3pt}}\kern-31.0pt\kern\wd2}

\def\verylongwick#1{\setbox2=\hbox{$\displaystyle#1$}
    \setbox3=\null\ht3=3.0pt\dp3=0.0pt\wd3=43.0pt
    #1\kern-\wd2\kern3.0pt\raise11.0pt\vbox{\hrule height0.3pt
    \hbox{\vrule width0.3pt\box3\vrule width0.3pt}}\kern-47.0pt\kern\wd2}

\definecolor{bluc}{cmyk}{1,1,0,0.1}
\definecolor{rossoCP3}{cmyk}{0,.88,.77,.40}
\definecolor{rosso}{cmyk}{0,1,1,0.4}
\definecolor{rossos}{cmyk}{0,1,1,0.55}
\definecolor{rossoc}{cmyk}{0,1,1,0.2}
\definecolor{verdes}{cmyk}{0.92,0,0.59,0.4}

\hypersetup{colorlinks, bookmarksopen, bookmarksnumbered,
citecolor=verdes, linkcolor=bluc, pdfstartview=FitH, urlcolor=rossos}

\theoremstyle{definition}

\theoremstyle{plain}

\def\lsim{\mathrel{\rlap{\lower4pt\hbox{\hskip1pt$\sim$}}
    \raise1pt\hbox{$<$}}}                % less than or approx. symbol
\def\gsim{\mathrel{\rlap{\lower4pt\hbox{\hskip1pt$\sim$}}
    \raise1pt\hbox{$>$}}}                % greater than or approx. symbol

\def\psibar{\overline{\psi}}

\def\eq#1{Eq.~(\ref{#1})}

\def\fig#1{Fig.~\ref{#1}}

\def\tab#1{Table~\ref{#1}}

\def\cite#1{\citep{#1}}

\newcommand{\mev}{\mathrm{MeV}}

\newcommand{\lra}{\longrightarrow}

\newcommand{\gv}{g_{\rm{VPP}}}
\newcommand{\mps}{m_{\rm{PS}}}
\newcommand{\fps}{F_{\rm{PS}}}

\newcommand{\mv}{m_{\rm{V}}}

\newcommand{\mpcac}{m_{\rm{PCAC}}}

\def\la{\langle}
\def\ra{\rangle}
\def\psibar{\overline{\psi}}

\def\tr#1{{\mathrm{tr}\left[#1\right]}}

\begin{document}

\title{Scattering of Goldstone Bosons and resonance production in a Composite Higgs model on the lattice}
\author[a]{Vincent   Drach}%\affiliation{\pu}\email{vincent.drach@plymouth.ac.uk}
\author[b]{Tadeusz  Janowski}%\affiliation{\uoe} 
\author[c]{Claudio Pica}%\affiliation{\cpthree}\email{pica@cp3-origins.net}
\author[d,e,f]{Sasa Prelovsek}%\affiliation{\ljubljana}\affiliation{ \ijs}\affiliation{ \ru}\email{sasa.prelovsek@ijs.si}

\affiliation[a]{Centre for Mathematical Sciences, Plymouth
  University, Plymouth, PL4 8AA, United Kingdom }
\affiliation[b]{School of Physics and Astronomy, University of Edinburgh, Edinburgh EH9 3FD, United Kingdom }
\affiliation[c]{CP3-Origins and eScience Center, University of Southern
  Denmark, Campusvej 55, DK-5230 Odense M, Denmark}
\affiliation[d]{Faculty of Mathematics and Physics, University of Ljubljana,  Slovenia}
 \affiliation[e]{Jozef Stefan Institute, Ljubljana, Slovenia}
\affiliation[f]{Institute for Theoretical Physics, University of Regensburg,   Germany}

\emailAdd{vincent.drach@plymouth.ac.uk}
%\emailAdd{vincent.drach@plymouth.ac.uk}
\emailAdd{pica@cp3-origins.net}
\emailAdd{sasa.prelovsek@ijs.si}

 %%%%%%%%%%%%%%%%%%%%%%%%%%%%%%%%%%%%%%%%%%%%%%%%%%%%%%%%%%%%%%%%%%%%%%%%%%%%%%%%%%%%%%%%
\abstract{
  \input{sections/abstract.tex}

}
\maketitle

%%%%%%%%%%%%%%%%%%%%%%%%%%%%%%%%%%%%%%%%%%%%%%%%%%%%%%%%%%%%%%%%%%%%%%%%%%%%%%%%%%%%%%%%
\section{Introduction}
\input{sections/introduction.tex}
%%%%%%%%%%%%%%%%%%%%%%%%%%%%%%%%%%%%%%%%%%%%%%%%%%%%%%%%%%%%%%%%%%%%%%%%%%%%%%%%%%%%%%%%
\section{Lattice setup}
\input{sections/lattice_setup.tex}

\section{Vector resonance from scattering of two pseudoscalars}
\input{sections/scattering.tex}

\section{Conclusions}
\input{sections/conclusions.tex}

%%%%%%%%%%%%%%%%%%%%%%%%%%%%%%%%%%%%%%%%%%%%%%%%%%%%%%%%%%%%%%%%%%%%%%%%%%%%%%%%%%%%%%%%

\section*{Acknowledgments}

We would like to thank D. Buarque Franzosi,  G. Cacciapaglia and  F.
Sannino   for useful discussions at various stages of this project.
The initial steps of this project have been performed on the HPC
facilities at the HPCC centre of the University of Plymouth. 
This work was performed using the Cambridge Service for Data Driven
Discovery (CSD3), part of which is operated by the University of
Cambridge Research Computing on behalf of the STFC DiRAC HPC Facility
(www.dirac.ac.uk). 
The DiRAC component of CSD3 was funded by BEIS capital funding via STFC capital grants ST/P002307/1 and ST/R002452/1
and STFC operations grant ST/R00689X/1. DiRAC is part of the National
e-Infrastructure. 
S.~P. was supported by
Slovenian Research Agency ARRS (research core funding No. P1-0035 and
No. J1-8137) and DFG grant No. SFB/TRR 55.  

\newpage
\input{sections/appendix.tex}

\vspace{0.5cm}

\newpage

%%%%%%%%%%%%%%%%%%%%%%%%%%%%%%%%%%%%%%%%%%%%%%%%%%%%%%%%%%%%%%%%%%%%%%%%%%%%%%%%%%%%%%%%
%\bibliography{paper_su2}

\bibliographystyle{JHEP} 
 \bibliography{paper_su2}

\end{document}

%% file: sections/abstract.tex
We calculate the coupling between a vector resonance and
two Goldstone bosons in $SU(2)$ gauge theory with $N_f=2$ Dirac fermions in
the fundamental representation.  The considered   theory  can be used to
construct a minimal Composite Higgs models. The coupling is related to the width of the vector resonance and we  determine it    by  simulating the scattering of two Goldstone bosons where the resonance is produced.  The  resulting coupling is $\gv=7.8\pm 0.6$, not far from $g_{\rho\pi \pi}\simeq 6$ in QCD.  This is the first  lattice calculation of the resonance properties   for a minimal
UV completion. 
This coupling controls the production cross section of the
lightest expected resonance at the LHC and enters into other tests of the Standard Model,   from Vector Boson Fusion to electroweak precision tests. Our prediction is crucial to constrain the model using lattice input and   for understanding the
behavior of the vector meson production cross section as a function of the
underlying gauge theory.   We also extract the  coupling $\gv^{\rm{KSRF}} =9.4 \pm 0.6$  assuming   the vector-dominance   and find that this phenomenological estimate slightly overestimates the value of the coupling.

%% file: sections/introduction.tex
%\item Why CH models 
%\item Resonances searches at the LHC
Among the numerous approaches to address the shortcomings of the
Standard Model, new strongly coupled sectors provide a
number of interesting mechanisms that address fundamental issues like
the naturalness problem or the origin of the Higgs' field. 

Pseudo Nambu-Goldstone Boson (PNGB) Composite Higgs models aim at identifying the Higgs
degrees of freedom with the Goldstone bosons of a new strongly coupled
sector \cite{Kaplan:1983fs,Kaplan:1983sm,Dugan:1984hq}. In this framework, the fundamental Higgs field of the Standard
Model is an effective low energy degree of
freedom at the electroweak scale  of a new strongly interacting sector
featuring spontaneous chiral symmetry breaking. The mechanism alleviates the
naturalness problem. The coupling of the new strongly interaction sector with the Standard
Model breaks explicitly the flavour symmetry of the underlying gauge
theory. This breaking is responsible for a non-trivial
potential for the pseudo-Goldstone bosons which provides the Higgs with a
mass and triggers electroweak symmetry breaking. % Additional interactions in the UV also contribute to the potential.

These models are constrained by comparing predictions at the electroweak scale
with the LHC data. Typically the predictions depends on low energy
constants (LECs) of the effective theory and on free parameters that
are related to the model building itself. %Lattice calculations provide estimates of the
%LECs and therefore reduce the dimensionality of the parameter space
%constrained by experiments. The lattice predictions can  be used to
%constrain models built on the same underlying gauge theory and
%serves as a laboratory to identify key differences and similarities among
%gauge theories used to build models of New Physics.
The vast majority of models are tested without specifying an underlying
strongly interacting sector, and by assuming that the LECs are free
parameters. While the approach allows to investigate entire classes of UV
completion, it neglects correlations among LECs and additional
information provided by a quantitative understanding of the strong dynamics.
The lattice approach allows to make first-principle predictions of the
low energy parameters and can  therefore provide stringent constraints
on some scenarios. 

Strongly interacting theories are expected to feature a rich spectrum
of resonances that modify the phenomenology at colliders. In minimal scenarios, vector resonance are expected to
mix with the electroweak bosons and can therefore be produced via
vector boson fusion or via the Drell-Yan production mechanism,
see for instance \cite{Contino:2015mha,Contino:2011np}.
These processes are controlled by the coupling $\gv$ of the vector
resonance ($V$) to two pseudoscalar Goldstone bosons ($P$).
 The phenomenology of new vector resonances is attracting considerable attention by
the community \cite{Gallinaro:2020cte,Liu:2019bua, Jamin:2019mqx,BuarqueFranzosi:2018eaj,Liu:2018hum,BuarqueFranzosi:2017ugz,Greco:2014aza}.  

In this work, we perform the first ab-initio
calculation of the  coupling  $\gv$ between a resonance and a pair of Goldstone bosons  using lattice techniques in
isolation of the Standard Model for a minimal
UV completion. We consider $SU(2)$ gauge theory with $N_f\!=\!2$ fundamental Dirac
fermions. The theory features an extended $SU(4)$ flavour symmetry
that spontaneously breaks to $Sp(4)$. It is used to build a PNGB Composite Higgs models in
\cite{Cacciapaglia:2014uja} and was recently reviewed in
\cite{Cacciapaglia:2020kgq}. In this model the physical Higgs boson is
a mixture of PNGBs and of the scalar state of the strong sector. The phenomenology of the  model   has been shown to be viable in view of the LHC
data in \cite{Arbey:2015exa}, and the mixing between the scalar
resonance and the Higgs can relax the bounds on the model
\cite{BuarqueFranzosi:2018eaj}. The phenomenology of vector
resonances for theories sharing the same chiral symmetry breaking
pattern is investigated in detail   as a function vector meson
coupling constant  in \cite{Franzosi:2016aoo}. The
authors  derived the bound from di-lepton and di-boson searches at the
LHC, as well as   the dependence of the electroweak precision
parameters as a function of the coupling constant. Our work can therefore
 be used to constrain a minimal model more efficiently.

We determine the $V\to PP$ coupling by extracting  the scattering amplitude for $PP\to V\to PP$ scattering in the vector channel via the rigorous  L\"uscher formalism \cite{Luscher:1990ux,
  Rummukainen:1995vs}. We apply    techniques  that have been widely used for $\pi\pi\to \rho \to \pi\pi $ scattering in QCD, for example \cite{Aoki:2007rd,Aoki:2011yj,Feng:2010es,Lang:2011mn,Dudek:2012xn,Erben:2019nmx,Alexandrou:2017mpi}.  The coupling $\gv$ is  inferred from a Breit Wigner parametrization of the
scattering amplitude. This work  represents  
 the first  fully-fledged scattering calculation of the coupling
$\gv$ in a 4D gauge theory with fermions beyond QCD.   
It complements previous studies addressing the phenomenology of
models based on the same underlying gauge theory \cite{Arthur:2016dir,Arthur:2016ozw,Arthur:2014zda,Drach:2015epq,Hietanen:2014xca,Hietanen:2013fya}. 

 The coupling  $\gv$  was previously  estimated via the phenomenological  relation $\gv^{\rm{KSRF}}=m_V/\fps$ by 
Kawarabayashi-Suzuki-Riazuddin-Fayyazuddin (KSRF)    \cite{Kawarabayashi:1966kd,Riazuddin:1966sw} that  assumes the vector-meson dominance.   These calculations were based on simulations where the vector state was stable.   
It was employed for QCD-like theories based on    $SU(3)$ \cite{Nogradi:2019auv,Nogradi:2019iek} as well as various groups $SU$ and  $Sp$ \cite{Nogradi:2019iek,Bennett:2019cxd,Bennett:2019jzz,
  Ayyar:2017qdf,Appelquist:2018yqe}. We refer to
\cite{Nogradi:2019auv} for a discussion of the estimation of
$\gv^{\rm{KSRF}} $ for a number of gauge theories.
 This relation turns out to be satisfied in QCD but has however never been
rigorously tested beyond before this work. It is therefore
crucial to guide model builders and searches for Beyond the Standard
Model Physics.

%% file: sections/lattice_setup.tex
%\item action
%\item phase diagram
%\item scale setting and Zs 

\label{sec:setup}

In the numerical simulations of  $SU(2)$ gauge theory with $N_f=2$, we choose a clover-improved Wilson action for the two flavours of Dirac fermions \cite{Wilson:1974sk,Sheikholeslami:1985ij} and the tree-level Symanzik improved action for the  gauge sector \cite{Luscher:1984xn}:  
\begin{align}
\label{lat_action}
    S &= -  \frac{\beta}{2} \sum_{x, \mu, \nu} c_0 \mathrm{Re Tr} P_{\mu \nu}(x) + c_1 \mathrm{Re Tr} \left( R_{\mu\nu}(x) + R_{\nu\mu}(x) \right)\\
      &+ \sum_{x,\mu} \bar \psi(x) \left( am_0 + 4 \right)\psi(x) - \frac{1}{2} \bar \psi(x+\mu) U_\mu (1 - \gamma^\mu ) \psi(x)~+~\frac{c_{sw}}{2} \sum_{x, \mu<\nu} \bar \psi(x) \sigma_{\mu\nu} \hat F^{\mu\nu} \psi(x).\nonumber
\end{align}
Here $U_\mu$ is the gauge field,  $\psi$ is the doublet of $u$ and $d$ fermions, $P_{\mu\nu}$ is the plaquette, $R_{\mu\nu}$ is a 2$\times$1 rectangular loop, $\hat F_{\mu\nu}$ is the Wilson clover term and $\beta$ is  the inverse lattice gauge coupling. The coefficient of the gauge action are set to $c_0 = 5/3$ and $c_0 + 8 c_1 = 1$  \cite{Luscher:1984xn}. The improvement coefficient $c_{SW}=1$ is set to its tree-level value in our simulations.  
The presence of bare mass term $ a m_0$ and the Wilson term explicitly breaks the $\mathrm{SU}(4)$ flavour symmetry to a $\mathrm{Sp}(4)$ subgroup. We use periodic boundary conditions in all four directions for the fermions.  

The results presented in this work are obtained at $\beta=1.45$. For each ensemble, we compute the pseudoscalar meson mass $\mps$, and  vector meson mass $\mv$ by fitting at large euclidean time the effective mass of appropriate two-points correlation function.  We also determine the bare quark mass $\mpcac$ defined through the Partially Conserved Axial Current relation and the bare pseudoscalar decay constant $\fps^{\rm{bare}}$, which renormalizes multiplicatively with the renormalization factor $Z_A$. The normalization of $\fps$ corresponds to the  convention where the pion decay constant in QCD is $130~\mev$. More details concerning our calculation of these quantities can be found in \cite{Arthur:2016dir} where we use the same strategy and normalization.

\begin{figure}[]
    \begin{center}
        \includegraphics[width=0.45\linewidth]{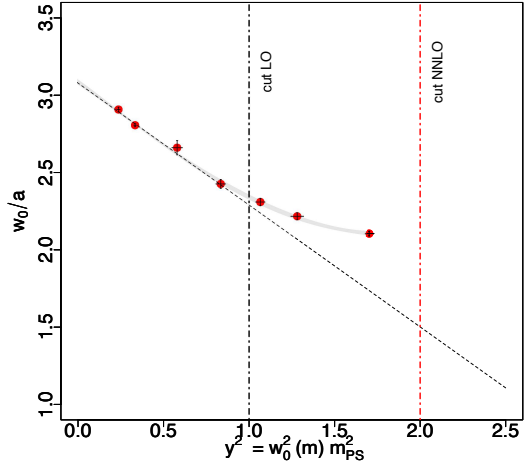} 
    \end{center}
    \caption{Chiral extrapolation of $w_0/a$ as a function of $y^2 = w_0(m) ^2 \mps^2$ .  The lattice data are in red. The LO fit is represented by a dotted straight line and the NNLO fit is represented by a shaded gray area. The vertical dotted line denotes the maximal value of $y^2$ included in the LO and NNLO fits.}\label{fig:WF}
\end{figure}

To compare our results with our previous results we convert the lattice quantities in physical units by setting the scale using the Wilson Flow observable $w_0$ \cite{Borsanyi:2012zs}. The scale $w_0$ is defined by 
$W(w_0^2) = W_{\mathrm{ref}}$, where $W(t) = t \tfrac{d}{dt} \left[t^2 E(t) \right]$ and $E(t)$\footnote{$E$ and $t$ are not to be confused with energy and Eucledian time employed later on.} is the flowed action density at flow time $t$.  The reference value $W_{\mathrm{ref}}$ is dimensionless and set to $1.0$, consistently with our previous definition. The value of $w_0/a$ is determined for a range of fermion masses and chirally extrapolated using NNLO expansion in terms of $\mps^2$ \cite{Bar:2013ora}. We employ ensembles with $\mps L >4 $, where the finite volume effects are small.  The data and are shown in \fig{fig:WF}, together with the best fitting curve, and the $1\sigma$ error band in gray. The final results reads $w_0^\chi/a~(\beta\!=\! 1.45)=3.08(2)$ using the NNLO fit which has a $\chi^2/ndof = 6.5/4$.

\begin{figure}[]
    \begin{center}
        \includegraphics[width=0.4\linewidth]{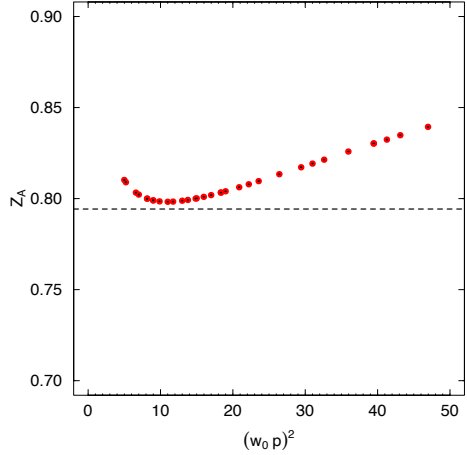}
    \end{center}
    \caption{Behaviour of $Z_A$ as a function of $(w_0 p)^2$ after the chiral extrapolation of $\Lambda_A$. The statistical errors are  included. The perturbative value  derived for the plaquette action and unimproved Wilson fermions   \cite{DelDebbio:2008wb} is shown by a black dashed line.} \label{fig:ZA} %
\end{figure}

In order to compare our result with the KSRF relation $\gv=m_V/\fps$, it is necessary to renormalize the pseudoscalar decay constant with $Z_A$. We performed the non perturbative determination of $Z_A$ in the RI'-MOM scheme \cite{Martinelli:1994ty}, using the same strategy as in our previous setup \cite{Arthur:2016dir}. The correlators are estimated using momentum sources introduced in \cite{Gockeler:1998ye}. To interpolate between the lattice momenta we furthermore use twisted boundary conditions \cite{Bedaque:2004kc,Sachrajda:2004mi}. The amputated vertex function  is then defined as follows: 
\begin{align}
\Pi_\mu(p)= S(p)^{-1}  \Big( \sum_{x,y,z} e^{-ip (x-z)} e^{-ip (z-y)} \la \psi(x)  \psibar(z)\gamma_{\mu}\gamma_5 \psi(z) \psibar(y) \ra \Big) S(p)^{-1}~,
\end{align}
where $S^{-1}(p)$ is the inverse propagator in spin and color space. Defining,
\begin{align}
\Lambda_A(p^2) &= -i \frac{\tr{\frac{\gamma_\mu \sin(ap_\mu)}{\sin^2 (ap_\mu)} S^{-1}(p)}}{\sum_\mu \tr{P_\mu \Pi_\mu(p)} }, \quad \textrm{with}\quad P_\mu = \frac{\gamma_{\mu}\gamma_5}{4}\,, 
\end{align}
it can be shown that in the chiral limit $\Lambda_A(p^2=\mu^2) \longrightarrow Z_A(\beta,\mu^2)$. The chiral extrapolation at fixed $p^2$ is performed using linear extrapolation in $\mpcac$ and our estimate of $\Lambda_A(p^2)$ in the chiral limit is shown in \fig{fig:ZA}.
We used  $(w_0^\chi p)^2 = 7 >(w_0^\chi m_V)^2 \sim 1 $ as a reference scale and our final estimate reads  $Z_A(\beta=1.45, (w_0^\chi p)^2 = 7) = 0.8022(3)$. The statistical errors are estimated using bootstrap resampling.

A range of simulations performed at $\beta=1.45$ allowed us to determine simulation parameters so that $\mv > 2 \mps$, therefore allowing the decay of the vector meson into two Goldstone bosons.
We used a single bare mass $m_0 = -0.6050$  to perform the calculation of the phase-shift of the two Goldstone boson scattering process, and two lattice volumes of $16^3\times 32$ and  $24^3\times48$. The value of the $\mps$, $\mpcac$ and $\mv^{\rm{naive}}$ are summarized in \tab{tab:lat}. The topological charge was monitored during the runs,  our simulations explore a range of topological sectors. In average the topological charge is compatible with zero.
The scattering study employs the quark propagators that are combined from the propagators with periodic and anti-periodic boundary conditions in time in order to effectively double the time extent of the lattice, as detailed for example in \cite{Lang:2011mn}. 
 
\begin{table}[h!]
\begin{tabular}{ccc|cc|cccc|} 
$\beta$ & $m_0$ & $V$ & Stat. & $\mps L$& $a\mpcac$ & $a\mps$ & $a\fps^{\rm{bare}}$ &  $am^{\rm{naive}}_V$  \\
\hline 
 $1.45$ & $-0.605$ & $16^3\times32$ & $1354$ & $3.6$ & $0.0107(7)$ &$0.226(4) $ & $0.048(2)$ & $0.488(42)$ \\
$1.45$ & $-0.605$ & $24^3\times48$ & $1857$ & $4.9$ &  $0.0107(2)$ & $0.205(2) $ & $0.058(1)$& $0.438(26)$ 
            \end{tabular}
\caption{   The main parameters of two lattice ensembles employed for the study of the vector resonance. The values obtained for $a\mpcac$, $a\mps$, $a\fps$ and $am^{\rm{naive}}_V$ are also provided.}\label{tab:lat}
\end{table}

%% file: sections/scattering.tex
The vector resonance  $V$ can decay into two pseudoscalar mesons $P$ as shown based on the symmetries in the next subsection.   The vector meson mass extracted from the two-point correlator on the chosen gauge ensemble is above the decay threshold $2\mps$ and the decay is therefore kinematically allowed. A resonance is not an eigenstate of the  Hamiltonian and its properties have to be inferred from the study of the scattering process $PP\to V\to PP$.  We will show that the investigation of this  resonance has similarities with the study of the  $\rho$-vector resonance   in QCD which has been  studied through $\pi\pi\to \rho\to\pi\pi$  in numerous lattice QCD investigations, for example \cite{Aoki:2007rd,Aoki:2011yj,Feng:2010es,Lang:2011mn,Dudek:2012xn,Alexandrou:2017mpi,Erben:2019nmx}. The main strategy is briefly reviewed here. The discrete eigen-energies $E_n$ of two pseudoscalars in the finite box of size $L$ and with periodic boundary conditions in space are determined. These are different from the non-interacting energies $E^{n.i.}=(\mps^2+{\mathbf p}_1^2)^{1/2}+(\mps^2+{\mathbf p}_2^2)^{1/2}$  of two pseudoscalars with momenta ${\mathbf p}_{1,2}=\tfrac{2\pi}{L}{\mathbf n}_{1,2}$ due to mutual interactions of pseudoscalars. The energy-shifts $\Delta E=E_n-E^{n.i}$ give information on their interactions and therefore phase-shift in partial wave $l$. L\"uscher's formalism   provides a rigorous relation between the eigen-energy $E_{CM}$ of two-particles in a finite box and their infinite-volume scattering amplitude $S_l(E_{CM})=e^{2i\delta_l(E_{CM})}$ at this energy $E_{CM}$ in center-of-momentum frame \cite{Luscher:1990ux,Luscher:1991cf}. We will extract  the phase shift for seven values of energy by considering the lowest two eigenstates $E_{n=1,2}$ in two different inertial frames and two different volumes\footnote{This renders eight   values of $E_{CM}$ and $\delta(E_{CM})$, but one of them  will have large uncertainty and will not be employed.}. 
 The energy-dependence of the phase shift will have a resonance shape. The resonance mass, width and the $V\to PP$ coupling $\gv$ will be extracted from this dependence via the Breit-Wigner type fit.  

\subsection{Goldstone Bosons scattering: flavour structure} 

We consider the scattering channels of the two-Goldstone boson scattering in $SU(2)$ gauge theory with $N_f=2$ fundamental fermions. The goal of this section is to show that one can employ the interpolators with the flavor structure analogous to QCD. For this purpose we demonstrate that the operators $\pi^- \pi^+ - \pi^+ \pi^- $ and $\rho_\mu$     belong to the  ten-dimensional irreducible representation of $Sp(4)$, where $\pi$ and $\rho$ are  bilinears
\begin{align}
\pi^+=  -\bar{d} \gamma_5 u,\quad \pi^- = \bar{u} \gamma_5 d , \quad \rho_\mu =  \bar{u} \gamma_\mu u-  \bar{d} \gamma_\mu d\,.
\end{align}
We follow the convention used for instance in \cite{Ryttov:2008xe} and summarized in appendix \ref{app:conv}, where 
 \begin{align}\label{eq:Q}
Q = \begin{pmatrix} 
u_L \\
d_L\\
\widetilde{u}_L\\
\widetilde{d}_L\\
\end{pmatrix} = \begin{pmatrix} 
u_L \\
d_L\\
(-i\sigma_2) C \bar{u}^T_R\\
(-i\sigma_2) C \bar{u}^T_R\\
\end{pmatrix}, \quad E = \begin{pmatrix} 0 & \mathbbm{1}_2 \\ -\mathbbm{1}_2  & 0 \end{pmatrix} \,,
\end{align}
the Pauli matrix acts in colour space and $C=i \gamma_0 \gamma_2$ is the conjugation charge matrix. It can be shown that the massless Lagrangian is symmetric under $SU(4)$ transformation, while the mass term can be shown to be $Sp(4)$ invariant.

 Let's  focus on  the multiplet    $\Pi^{i=1,\dots,5}$ and the decoupled $V^{a=1,..,10}_{\mu}$  defined as  
\begin{align}\label{eq:GBs}
\Pi^i =Q^T (-i\sigma_2) C \gamma_5 X^i E Q + \left(Q^T (-i\sigma_2) C \gamma_5 X^i E Q\right)^\dagger\ , \qquad V^a_{\mu} = \overline{Q} S^a \gamma_\mu Q ~. 
\end{align}
Here   $X^{i=1,\dots,5}$ are the broken generators used to parametrized the coset $SU(4)/Sp(4) $, while  $S^{a=1,\dots,10}$ are the generators of $Sp(4) $ defined in the appendix.

  Consider an infinitesimal the infinitesimal transformation  $Q \lra \left(\mathbbm{1}_4 + i \alpha^a S^a \right) Q$ 
where $\alpha^{i=1,\dots,10}$ are real parameters. The explicit calculation shows that the multiplet $\bf{\Pi}$   transforms  as  ${\bf{\Pi}} \lra  {\bf{\Pi}} + G {\bf{\Pi} }$, where $G = 2 i\sqrt{2} \alpha^a T^a $ is a skew-symmetric matrix which can be written as a linear combination of the generators $T^{a=1,\dots,10}$  of $SO(5)$  defined in the appendix. The multiplet $\bf{\Pi}$ therefore transforms as a $5-$dimensional irreducible representation of $Sp(4)$.

The transformation of the tensor product of two Goldstone Bosons operators ${\bf \Pi} \otimes {\bf \Pi} $ can then be  worked out. In this work, we focus on the antisymmetric part of the tensor:
\begin{align}
H^{ij} = \frac{1}{2}\left( \Pi^i\Pi^j - \Pi^i\Pi^j\right)\,.
\end{align}
  Defining the $10-$dimensional vector $F^a = \tr{T^a  H}$,  an explicit calculation shows that $F^a \lra F^a + 2i f_{abc} \alpha^b F^c$, 
where $f_{abc}$ are the structure constant of $Sp(4)$. The multiplet $F^{a=1,\dots,10}$ therefore belongs to a $10-$dimensional irreducible representation of $Sp(4)$. 
 Furthermore, expressing \eq{eq:GBs} in terms of the $u$ and $d$ fields via \eq{eq:Q}, using properties of the conjugation charge matrix and of the skew-symmetric matrix $(-i \sigma_2)$ acting in color space, we find that $F^3 = \tfrac{1}{4} \left( \pi^- \pi^+ - \pi^+ \pi^- \right)$.  In a similar manner    the bilinears $V^a_{\mu}$ transform as   $V^a \lra V^a + i f_{abc}  \alpha^b S^c$ and   $V^3_{\mu} = \rho_\mu$. 
  %\label{eq:I0pipi} 

%We show in this section that the field bilinear:
%\begin{align}%
%O_\mu(x) =  \bar{u} \gamma_\mu u(x) -  \bar{d} \gamma_\mu d(x)
%\end{align}
%belongs to a $10$ dimensional irrep of $Sp(4)$. Introduce $S^{a=1,\dots,10}$ a basis of the Lie algebra $sp(4)$ and the corresponding structure constant $f_{abc}$ as defined in the appendix\ref{appendix}, and consider 

In other words, we have shown that the usual $3$-dimensional irrep ($I=1$) of the isospin group is contained in the $10$-dimensional representation of $Sp(4)$. Regarding discrete symmetries, it straightforward to check that $F^3$ and $ \rho_\mu$ have negative parity and 
negative charge conjugation.

\subsection{Operators and correlators}
The aim is 
to extract eigen-energies of the system with the  quantum numbers  of the vector resonance, namely $J^P=1^-$.  Given the flavour structure discussed in the previous section, we  employ the two following operators:
 \begin{align}
  \label{O}
    O_{PP}(t, \mathbf p, \mathbf  0) &\!=\! O_1\! =\! \frac{1}{\sqrt{2}}\sum_{\mathbf x} \bar d(x) \gamma^5 u(x) e^{i \mathbf p \cdot \mathbf x}  \sum_y \bar u(y)\gamma^5 d(y) e^{i \mathbf y \cdot \mathbf 0}-\{u \leftrightarrow d\} \propto P^+(\mathbf p)P^-(0)\!-\!P^-(\mathbf p)P^+(0)\nonumber\\
    O_V(t, \mathbf p) &\!=\! O_2\!=\! \frac{1}{\sqrt{2}}\sum_{\mathbf x} \bar u(x) (\gamma \cdot \tfrac{\mathbf{p}}{|\mathbf{p}|})  u(x) e^{i \mathbf p \cdot \mathbf x}~-~\{u \leftrightarrow d\}~.
  \end{align}
To compute the phase shift for different values of $E_{CM}$ we consider two inertial frames with total momenta ${\mathbf P_{tot}}={\mathbf p}=(0,0,1)$ and $(1,1,0)$ in units of $2\pi/L$. The operators   with momenta $ (0,0,1)$  transform according to the irreducible representation $A_2^-$ of the group  $D_{4h}$ within the notation of \cite{Rummukainen:1995vs}. The operators   with momenta $(1,1,0)$ transform according to $B_1^-$ of the group  $D_{2h}$ in the notation of \cite{Feng:2010es}.

 The eigen-energies are obtained by computing in each frame the  $2\times 2$ correlation matrix
\begin{align}\label{eq:C}
C(t,\mathbf p) &= \begin{pmatrix} \langle O_{PP}(t, \mathbf p, \mathbf  0)  O^\dagger_{PP}(0, \mathbf p, \mathbf  0)\rangle & \langle O_{PP}(t, \mathbf p, \mathbf  0)  O^\dagger_{V}(0, \mathbf p) \rangle \\
\langle O_{V}(t, \mathbf p)  O^\dagger_{PP}(0, \mathbf p, \mathbf  0) \rangle &   \langle  O_V(t, \mathbf p)  O^\dagger_{V}(0, \mathbf p, \mathbf  0) \rangle \end{pmatrix}~,
\end{align}  
which renders the Wick contractions analogous to those in QCD. We evaluate those with the  stochastic $U(1)$ noise technique used  in simulations \cite{Aoki:2007rd,Aoki:2011yj}  of $\rho$-meson decay    in QCD. 

The experience from analogous studies of  $\pi\pi \to \rho \to \pi \pi$ scattering in QCD indicates that  it is crucial to employ  one  $\bar{q} q$ operator and the $\pi \pi$ operators relevant in the given energy range to obtain eigen-energies of the system. 
Adding more  $\bar{q} q$ operators with given quantum numbers does not affect the eigen-energies, as illustrated for example in Fig. 8 of \cite{Lang:2011mn}. 
Therefore we expect that two employed operators suffice to extract eigen-energies in this energy range.

 \subsection{Eigen-energies}\label{sec:En}

The time-dependence of the correlation matrix $ C_{ij}(t)= \langle 0|O_i (t)O_j^\dagger (0)|0\rangle=\sum_n \langle 0|O_i|n\rangle e^{-E_nt} \langle n|O_j^\dagger|0\rangle$ contains information on the eigen-energies $E_n$. We extract those using the widely used GEVP variational method \cite{Michael:1985ne,Luscher:1990ck,Blossier:2009kd}
\begin{equation}
C(t)u^{(n)}(t)=\lambda_n(t)C(t_0)u^{(n)}(t_0)\ , \quad \lambda_n(t)\overset{\mathrm{large}~ t}{\longrightarrow} A_n ~e^{-E_n t}
\label{gevp}
\end{equation}
 with reference time $t_0=4$.

  \begin{figure}  
  \includegraphics[width=0.45\textwidth]{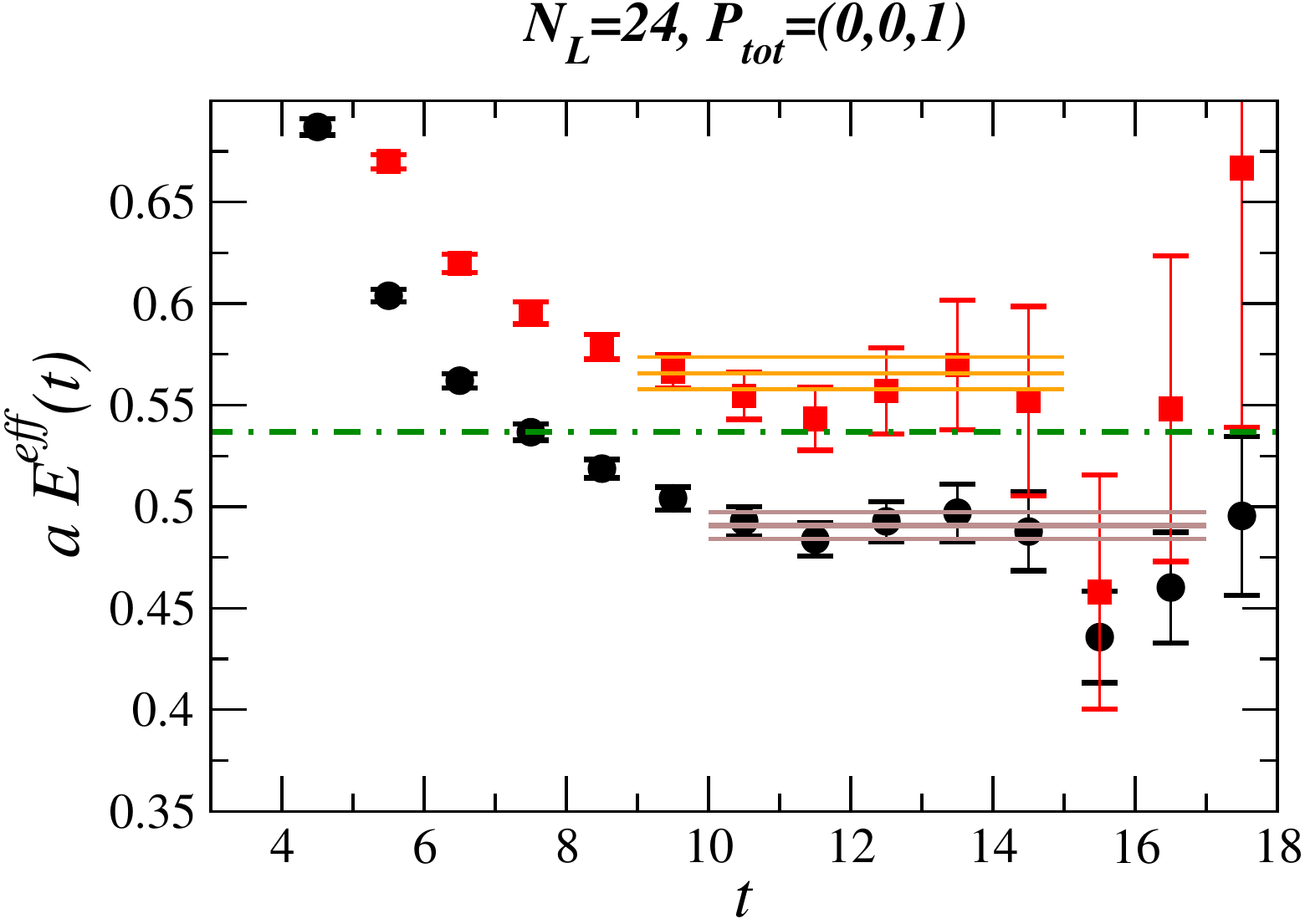}\quad \includegraphics[width=0.45\textwidth]{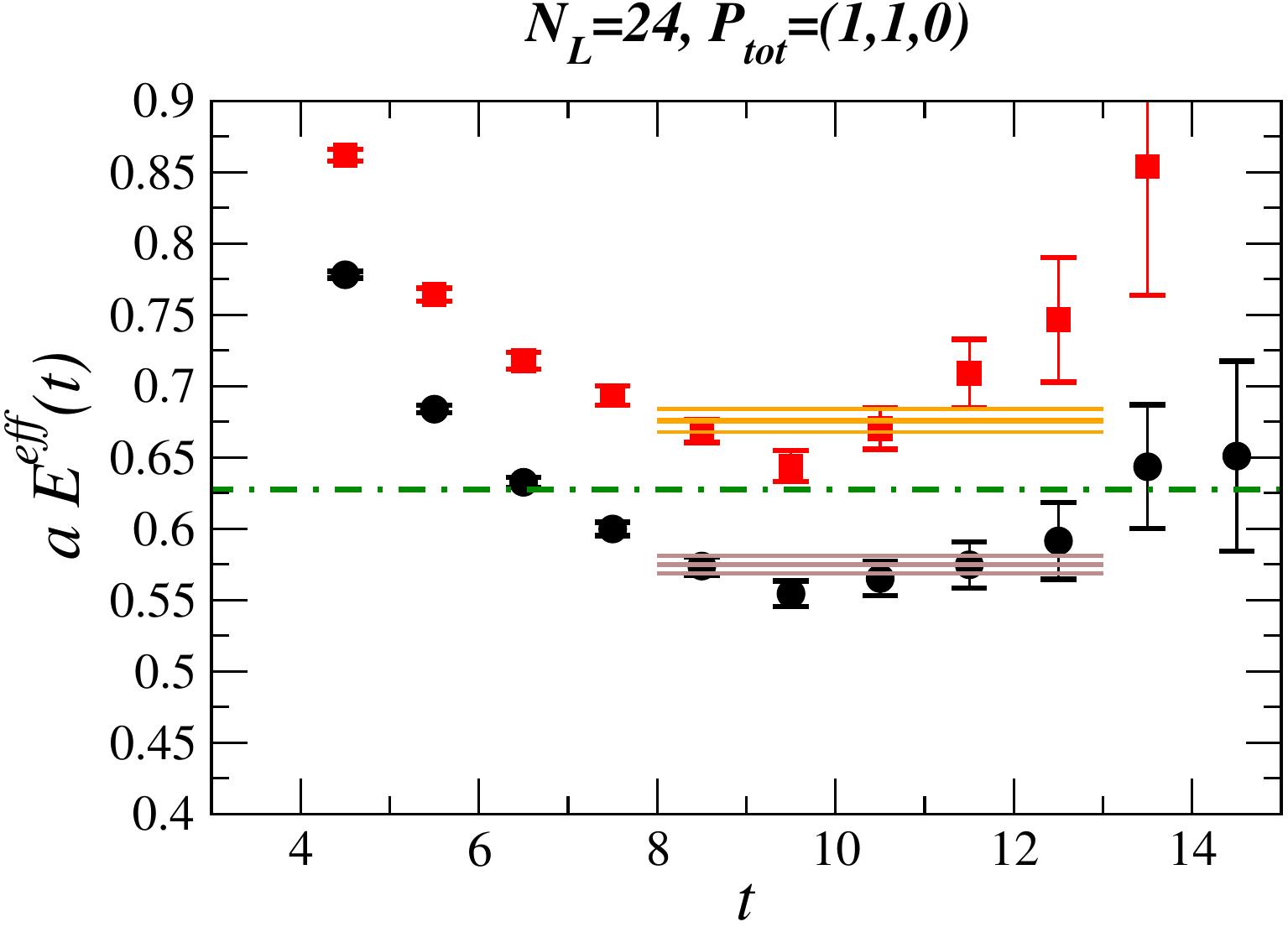}
\includegraphics[width=0.45\textwidth]{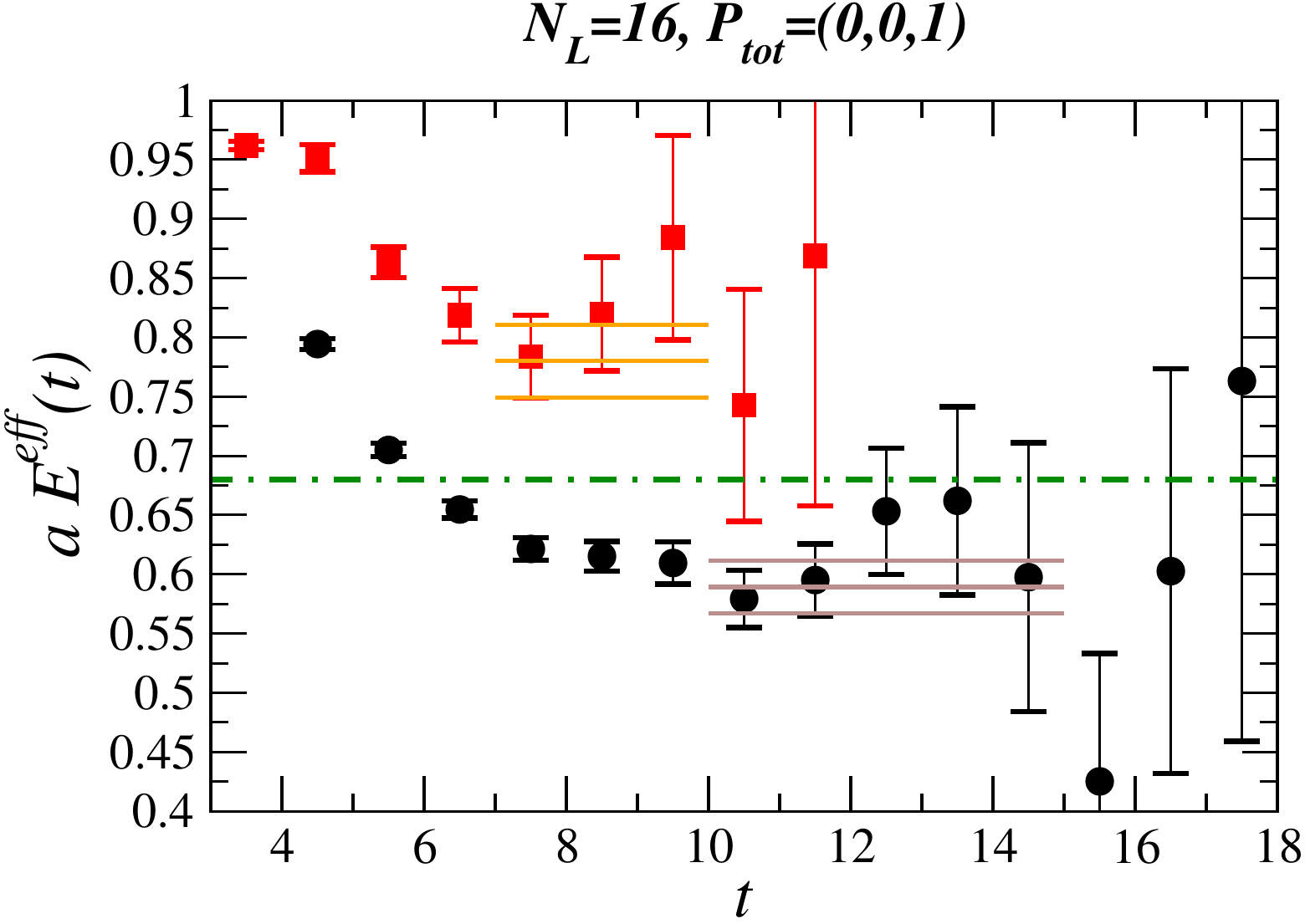}\quad \includegraphics[width=0.45\textwidth]{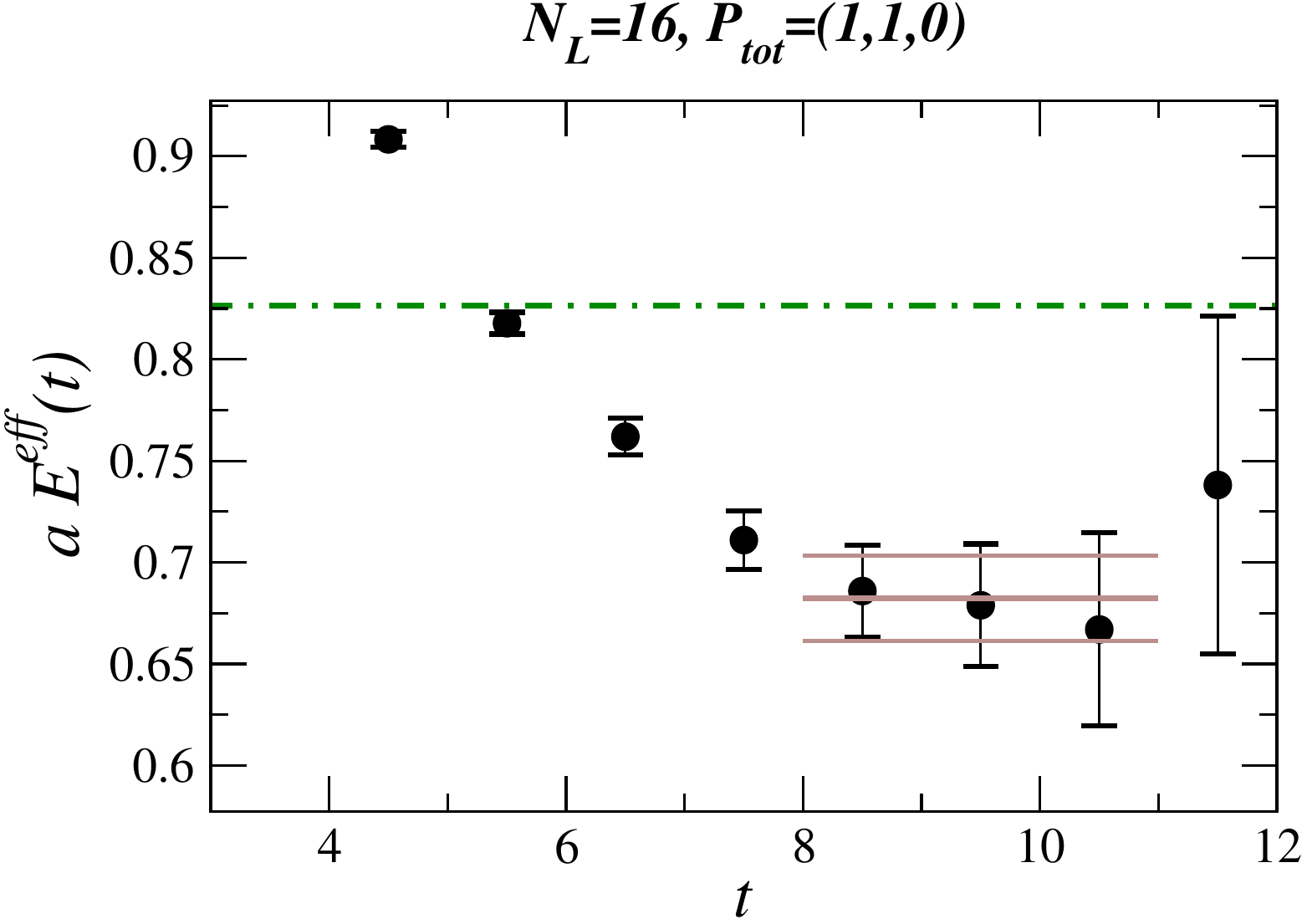}
\caption{ The effective energies  $E_n^{eff}(t)$ for seven eigenstates of the two-pseudoscalar system on a finite lattice.   The fitted eigen-energies $E_n$   are also provided together with the fit-ranges (\ref{gevp}). The dot-dashed lines indicate non-interacting energy $E^{n.i.}=(\mps^2+{\mathbf p}^2)^{1/2} +\mps$  of two -pseudoscalar system $P(0)P({\mathbf p})$.  } \label{fig:Eeff}
\end{figure}

  The effective energies $E_n^{eff}(t)=\log[\lambda_n(t)/\lambda_n(t+1)]$ for seven eigenstates are given in Fig. \ref{fig:Eeff} and  they are related to  $E_n$ in the plateau region. They correspond to lowest two eigenstates for two inertial frames and two volumes; the first excited state  for $N_L=16$ and ${\mathbf P}_{tot}=(1,1,0)$ has a large statistical uncertainty and will not be used in analysis.  
   We do not find an indication for non-exponential time-dependence at large $t$ before the signal is lost in the noise since the   size of the time-direction has been effectively doubled by combining periodic and antiperiodic propagators in time. 
      The eigen-energies $E_n$ are extracted from the correlated one-exponential fits (\ref{gevp}) of eigenvalues $\lambda_n(t)$ in the plateau region, that are indicated in the Fig. \ref{fig:Eeff} and tabulated in \ref{tab:En}.  
       
 \begin{table}[h!]
\begin{tabular}{ccc|cccccc}
     $N_L$ &  ${\mathbf P}_{tot}$  & level $n$  & fit-range &  $E_na$ & $E_{CM}a$ & $(\tfrac{p_*L}{2\pi})^2$ &  $\tfrac{(p_*a)^3 \cot\delta_1}{E_{CM}a}$ & $\delta_{l=1}[^\circ ]$     \\
     \hline     
    24  & (0,0,1)  & 1 & 10-17 & 0.491(7) &   0.415(8) & 0.016(23)  & 0.0055(11) &  0.9(1.4)  \\
      24 & (0,0,1) & 2 & 9-15   & 0.566(8) &   0.501(9) & 0.30(3)      &  -0.027(5) & 167(4)\\
       24 & (1,1,0) & 1 & 8-13  & 0.575(6)  &   0.440(8) & 0.093(27)  &  0.0045(11) & 14.3(2.2)\\
        24 & (1,1,0) & 2 &8-13 &  0.676(8) &    0.565(9) & 0.55(4)     &   -0.025(3) & 152(4.9)\\
        
       16   & (0,0,1) & 1 & 10-15& 0.589(22) & 0.440(30) & -0.020(42) & 0.013(3) &  -1.7(5.8)i\\
       16    & (0,0,1) & 2 & 7-10 & 0.780(31) & 0.674(36) &  0.40(8)    & -0.044(8) & 152(10)\\
        16    & (1,1,0) & 1 & 8-11 & 0.682(21) & 0.397(36)  & -0.078(46) & 0.0085(17) & -24(20)i\\ 
      \hline 
            \end{tabular}
\caption{Seven eigen-states of $PP$ system   and the resulting information related to the scattering phase shift $\delta_l$ for partial-wave $l=1$.   }\label{tab:En}
\end{table}
 
\subsection{Scattering amplitude and  phase shift}

The eigen-energies $E_n$ in Fig. \ref{fig:Eeff} (solid lines) are  shifted with respect to non-interacting energy $E^{n.i.}=(\mps^2+{\mathbf p}^2)^{1/2} +\mps$  of two-pseudoscalar system $P({\mathbf p})P(0)$ (dot-dashed lines). These non-zero shifts are essential for extracting the scattering information. 

The relation between eigen-energy $E_{CM}$ of two particles  and infinite-volume scattering phase shift $\delta(E_{CM})$ at that energy was derived by L\"uscher  \cite{Luscher:1990ux} for  ${\mathbf P}_{tot}=0$. This relation was generalized  in \cite{Rummukainen:1995vs,Feng:2010es} for inertial frames with ${\mathbf P}_{tot}=(0,0,1)$ and $(1,1,0)$ that are employed here.   In this case, the eigenstate with energy $E_n$ corresponds to  the energy  $E_{CM}=(E_n^2-{\mathbf P}_{tot}^2)^{1/2}$ in the center-of-momentum frame, where   pseudoscalars $P(\mathbf{p}_*)P(-\mathbf{p}_*)$ have back-to-back momenta $p_*= (E_{CM}^2/4-\mps^2)^{1/2}$ and dimension-less momenta $q= p_*L/(2\pi)$. The relation between eigen-energy (or corresponding $q$) and the scattering phase-shift   at that energy  (or   $q$) is given by  
\begin{align}
\label{luscher-type}
\tan \delta_1(q)&=\frac{\pi^{3/2}q\gamma}{Z_{00}(1;q^2) + \frac{2}{\sqrt 5 q^2} Z_{20}(1;q^2)}\quad \mathrm{for}\ \mathbf P_{tot}=(0,0,1) \\
\tan \delta_1(q)&= \frac{\pi^{3/2}q\gamma}{Z_{00}(1;q^2) - \frac{1}{\sqrt 5 q^2} Z_{20}(1;q^2) -i\frac{\sqrt{3}}{\sqrt{10}q^2}\left( Z_{22}(1;q^2) - Z_{2(-2)}(1;q^2) \right)}\quad \mathrm{for}\  \mathbf P_{tot}=(1,1,0)~, \nonumber
\end{align}
for the relevant irreducible representations of the symmetry groups listed after Eq. (\ref{O}). 
Here  $\delta_1$ denotes phase shift for partial-wave $l\!=\!1$,   $ Z_{lm}(s,q^2) = \sum_{n\in P_d}Y_{lm}(n)/(q^2-n^2)^s$ \cite{Feng:2010es}  and $\gamma=(1-v^2)^{-1/2}$. 

\begin{figure} 
\includegraphics[width=0.49\textwidth]{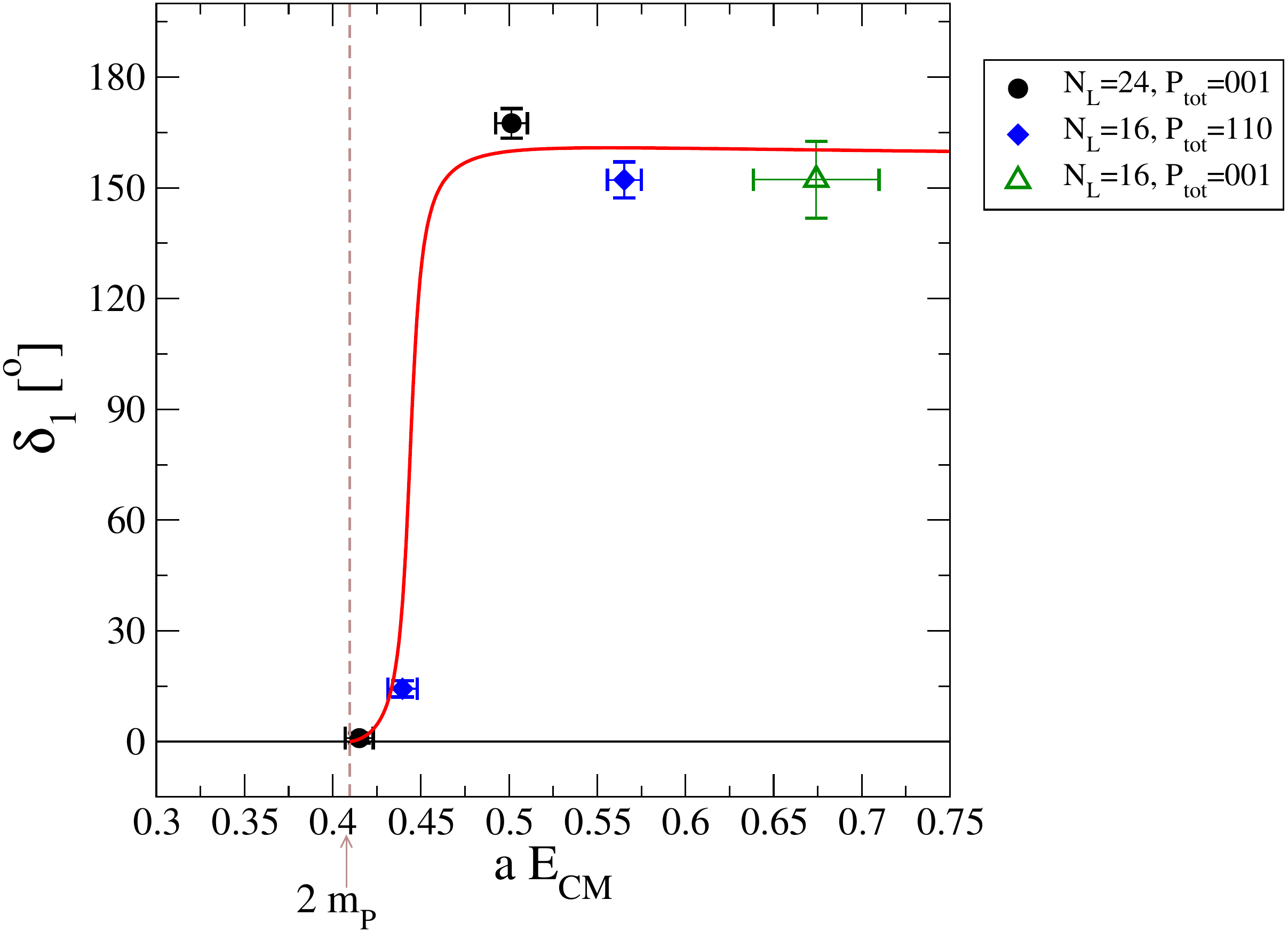}$\ $  
\includegraphics[width=0.49\textwidth]{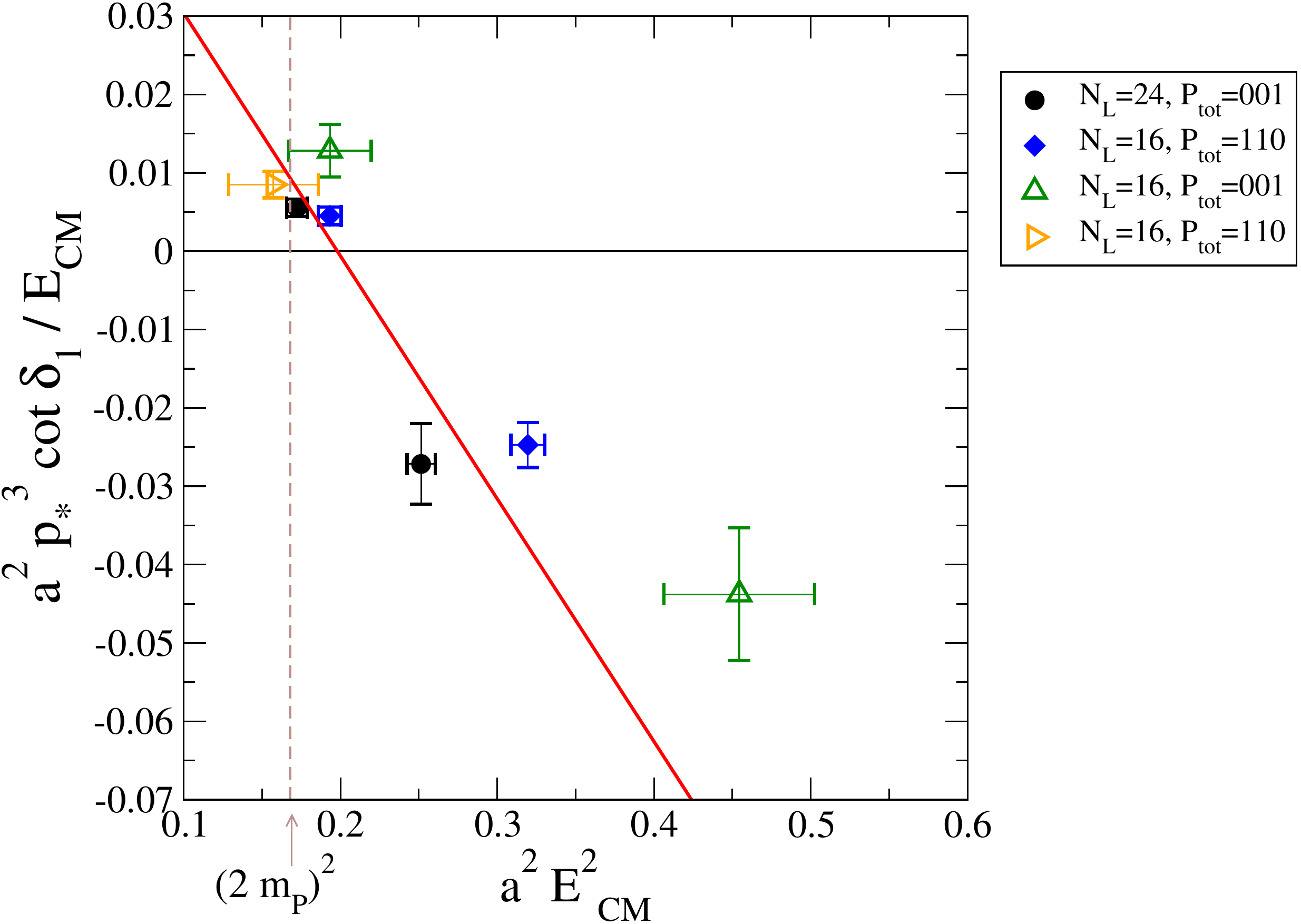}  
\caption{   Results on the scattering $PP$ of two pseudoscalars in partial-wave $l\!=\!1$. (a) 
Scattering phase-shift $\delta_1$ as a function of center-of-momentum energy $E_{CM}$.   (b) Quantity $p_*^3 \cot\delta_1(E_{CM})/E_{CM}$ as a function of $E_{CM}^2$, which is expected to be linear in case of Breit-Wigner resonance. The red solid line shows the Breit-Wigner resonance fit (\ref{bw1},\ref{p3cot})  with the values of resonance parameters (\ref{res_par}). Dashed vertical line indicates the position of the threshold for $N_L=24$. The left plot omits the points below threshold that lead to imaginary phase shifts. } \label{fig:delta}
\end{figure}

Every eigen-energy $E_n$  renders certain $E_{CM}$, $q$  and   phase shift, which are tabulated for seven eigenstates in Table \ref{tab:En}. The dependence of the phase-shift on  $E_{CM}$ is plotted in Fig. \ref{fig:delta}a.  It increases from the small values, past the resonance value $90^\circ$ to the large values.

\subsection{Mass $m_R$, width $\Gamma$ and the coupling $\gv$   of the vector resonance}

  The energy dependence of the   phase shift for $PP$ scattering in Fig. \ref{fig:delta}a shows a behavior expected for a narrow resonance that lies slightly above threshold: it rises from small values, through the resonance value $90^\circ$ at about $m_Ra=E_{CM}a\simeq 0.45$,      to the large   values close to $180^\circ$.  It is expected to have a Breit-Wigner type resonance form in the vicinity of a narrow vector resonance  
    \begin{equation}
    \tan \delta_1(E_{CM}) =  \frac{E_{CM}~\Gamma(E_{CM})}{m_R^2-E_{CM}^2}, \qquad   \Gamma(E_{CM})= \frac{\gv^2}{6\pi}\frac{p_*^3}{E_{CM}^2}~.
    \label{bw1}
  \end{equation}
 The resonance width $\Gamma(E_{CM})$ is parametrized in terms of the $V\to PP$ coupling $\gv$  and  the phase space for partial wave $l=1$, where $p_*$ is the momentum of the pseudoscalar in CM frame.  The resonance phase shift (\ref{bw1}) implies that  $p_*^3 \cot \delta_1/E_{CM} $ is a linear  function of $E_{CM}^2$
  \begin{equation}
        \frac{p_*^3 ~\cot \delta_1(E_{CM})}{E_{CM}} = \frac{6 \pi}{\gv^2} \left( m_R^2 - E_{CM}^2 \right)~.
          \label{p3cot}
  \end{equation}
  This quantity is shown  in Fig.  \ref{fig:delta}b for $PP$ scattering from our lattice simulation. It is falling with energy, it crosses through the resonance value  $\cot \delta_1=0$  at $E_{CM}=m_R$ and   roughly supports the linear dependence on $E_{CM}^2$.  

Finally, we proceed to  determine  the vector resonance parameters from the phase shift that  shows a resonant shape. We aim to determine the mass $m_R$ and the coupling $\gv$  that parametrizes the width rather than the width itself. The width  is   strongly dependent on  $\mps$ through the phase space, while the dependence of  $\gv$ on $\mps$ is expected to be much milder  (the dependence of   $g_{\rho\pi\pi}$  on $m_\pi$ in QCD   is very mild as evidenced in the review of various lattice results  \cite{Alexandrou:2017mpi,Erben:2019nmx}).
The $m_R$ and $\gv$ can be  read-off from zero and the slope for the quantity (\ref{p3cot}) in Fig. \ref{fig:delta}b.  In order to take the correlations  properly into account, we extract $m_R$ and $\gv$ by minimizing the correlated $\chi^2$
\begin{equation}
\chi^2=\sum_{k=1}^{N_k}\sum_{k^\prime=1}^{N_k} [E_k-E_k^{par}(m_R,\gv)]~\mathrm{ Cov}^{-1}(k,k^\prime)~[E_{k^\prime}-E_{k^\prime}^{par}(m_R,\gv)] 
\label{chisq}
\end{equation}
with $N_k=7$ for seven eigenstates. 
Here  $E_{k=1,..,N_k}$ are energies of    seven eigenstates determined on the lattice. $\mathrm{Cov}(k,k^\prime)$ is their $7\times 7$ covariance matrix, where  $\mathrm{Cov}(k,k^\prime)=0$ if eigenstates $k$ and $k^\prime$ correspond to ensembles with different $N_L$. The $E_k^{par}(m_R,\gv)$ is the   value of $E_k$ determined analytically  for given $m_R$ and $\gv$ based on the parametrization (\ref{p3cot}) and $\delta_1$ from L\"uscher-type relation (\ref{luscher-type}).  

The final   parameters of the vector resonance based on minimization of $\chi^2$ (\ref{chisq}) are
\begin{align}
m_Ra&=0.445\pm 0.008 \pm 0.002 \ \quad  \gv=7.8\pm 0.5 \pm 0.1 \nonumber\\
&\mathrm{with} \quad \mathrm{cor}(m_R,\gv)=0.35 \quad \mathrm{and} \quad \chi^2/\mathrm{dof} =1.5~\ .
\label{res_par}
\end{align} 
The first error is statistical and  arises from the errors on the energies  encoded in the covariance matrix, while the second error  accounts for the uncertainty in the pseudoscalar mass.    The correlation between resonance parameters  is provided by 
$\mathrm{cor}\!=\!\frac{\langle (g-\langle g\rangle )(m_R-\langle m_R\rangle )\rangle}{\sqrt{\langle (g-\langle g\rangle )^2\rangle}   \sqrt{\langle (m_R-\langle m_R\rangle )^2\rangle}}$.  The phase-shift dependence based on these resonance parameters (\ref{res_par}) is plotted by red solid lines in Fig. \ref{fig:delta}.
  The results (\ref{res_par}) are  based on seven eigenstates from both volumes $N_L=24,16$. Restricting the fit to four eigenstates on the larger volume  $N_L=24$, one gets compatible results $m_Ra=0.442\pm 0.009\pm 0.002, \    \gv=7.7\pm 0.5\pm 0.1$ with $ \mathrm{cor} =0.37$ and $\chi^2/dof=1.7$. For comparison, Table \ref{tab:lat} presents the mass of a vector particle ($am^{\rm{naive}}_V$) obtained using 
a single $\bar{q}q$ operator in a conventional way.

Note that the calculation is performed at the finite pseudoscalar meson mass and at finite lattice spacing. However it should be noted that lattice  results on $g_{\rho\pi\pi} $ in QCD do not show large discretization errors  and that its dependence on $m_\pi$  is very mild: the compilation   \cite{Alexandrou:2017mpi,Erben:2019nmx} of lattice  simulations by several groups shows that the coupling is $g_{\rho\pi\pi}\simeq 6$ for a wide range of $m_\pi$ and lattice spacings.  
 We therefore believe our calculation provides a first estimate of the coupling $\gv$ in the chiral limit of $SU(2)$ gauge theory with two fundamental flavors of Dirac fermions.

\subsection{Comparison of the coupling $\gv$ with the KSRF relation}

The $V\to PP$ coupling is often determined phenomenologically via the   KSRF relation $\gv^{\rm{KSRF}}=\mv/\fps$    based on the vector-meson dominance \cite{Kawarabayashi:1966kd,Riazuddin:1966sw}. This relation gives us 
\begin{equation}
\gv^{\rm{KSRF}} =m_V^{naive}/(Z_A F_{PS}^{bare})=9.4(6)
\end{equation}
 at finite $\mps$, $V=24^3\times48$  and $Z_A$ provided in Section \ref{sec:setup}. The central  value $\gv^{\rm{KSRF}}$  is   $20\%$ larger than  the value $\gv$ (\ref{res_par})  rigorously  extracted from the scattering. 

Note furthermore that using our previous calculation with an unimproved action \cite{Arthur:2016dir}, and using the value of $\mv$ and $\fps$ extrapolated to the chiral and continuum limit, we found $\gv^{\rm{KSRF}} =  13.1(2.2)$. The value $\gv^{\rm{KSRF}} = 15.6(4)$  used in \cite{Nogradi:2019auv}   is  from an update analysis including more gauge ensembles \cite{Drach:2017btk}.
Our calculation therefore suggests that the KSRF relation overestimates the value of the coupling $\gv$ for $SU(2)$ gauge theory with two fundamental flavors of Dirac fermions and that the control of systematics must be improved in order to clarify the overal picture as a function of the number of flavours.

%% file: sections/conclusions.tex
{We  determined the coupling $\gv$ between a vector resonance and a pair of Goldstone bosons in a minimal realization of a composite Higgs model.   This is the first lattice  result for the coupling of a decaying resonance  in a such 4-dimensional gauge theories with fermions.
The calculation
is performed by considering the scattering of two Goldstone bosons in the channel of the vector resonance.
Assuming that the chiral dependence of $\gv$ is mild (like in QCD) and
that the discretization error are negligible, we  obtain a  rigorous estimate of
the coupling in the continuum and in the chiral limit based on the non-perturbative calculation. 
The result suggests a rather large value of
the coupling but smaller than the one obtained using the KSRF
relation. Additional numerical simulations would be required to obtain a more
definite answer.
The coupling is relevant to constrain the phenomenology of a vector
resonance in composite Higgs model at the LHC, and bring additional
constraints on Vector Boson Scattering and on electroweak precision
tests.

%% file: sections/appendix.tex
\begin{appendix}
\section{Conventions}\label{app:conv}

We follow the convention used in \cite{Ryttov:2008xe}, where 
\begin{displaymath}
\sigma_1 = 
\left(\begin{array}{c c c}
0 & \phantom{0} & \phantom{-}1 \\
1 & \phantom{0} & \phantom{-}0
\end{array}\right), \quad
\sigma_2 = 
\left(\begin{array}{c c c}
0 & \phantom{0} &-i \\
i & \phantom{0} & 0
\end{array}\right), \quad
\sigma_3 = 
\left(\begin{array}{c c c}
1 & \phantom{0} & 0 \\
0 & \phantom{0} & -1
\end{array}\right),
\quad
\sigma_4 = 
\left(\begin{array}{c c c}
1 & \phantom{0} & 0 \\
0 & \phantom{0} & 1
\end{array}\right), 
\quad
E = 
\left(\begin{array}{c c c}
0 & \phantom{0} & \sigma_4 \\
-\sigma_4 & \phantom{0} & 0
\end{array}\right) 
\end{displaymath} 
and 
\begin{align}
B_1 =  \sigma_4, B_2 =  i \sigma_4, B_3 =  \sigma_3, B_4 =  i \sigma_3, B_5 =  \sigma_1, B_6 =  i \sigma_1, \quad  D_4 = \sigma_2, D_5 = i\sigma_2\, .\nonumber
\end{align}
 The fifteen generators of $SU(4)$ are denoted $S^{a}$ and $X^i$ with $a=1,\dots,10$ and $i = 1,\dots5$.
The generators $S^a$ satisfy the relation $(S^a)^T E + E S^a = 0 $ and are a representation of $Sp(4)$: \begin{align}
S^{a} &= \frac{1}{2\sqrt{2}} \begin{pmatrix} \sigma_i &0 \\
0 & -\sigma^T_i
\end{pmatrix},  \quad a = 1,\dots, 4 \qquad 
S^a = \frac{1}{2\sqrt{2}}\begin{pmatrix} 0 & B_{i} \\
 B_{i}^\dagger& 0
\end{pmatrix}, \quad a = 5,\dots, 10 \nonumber\\
  X^i  &= \frac{1}{2\sqrt{2}} \begin{pmatrix} \sigma_i &0 \\
0 & \sigma^T_i
\end{pmatrix},  \quad i = 1,\dots, 3 \qquad \  \ 
X^i = \frac{1}{2\sqrt{2}} \begin{pmatrix} 0 &D_i \\
 D_i^\dagger& 0
\end{pmatrix}, \quad i = 4, 5~. \nonumber
\end{align}
The generators are normalized so that :
\begin{align}
\tr{ S^a S^b } = \frac{1}{2} \delta^{ab}, \quad \tr{ X^i X^j } = \frac{1}{2} \delta^{ij}, \quad \tr{ S^a X^i } = 0 \,.\nonumber
\end{align}
The structure constant of the algebra of $Sp(4)$ are defined as $f_{abc} = 2 \tr{ S^a [S^b,S^c]}$.

The 10 generators of $SO(5)$ with normalization $Tr[T^i T^j] = \frac{1}{2} \delta^{ij}$ are defined as follows:
\begin{align}
T^1 &= \frac{i}{2} \begin{pmatrix}
 0 & 0 & 0 & 0 & 0 \\
 0 & 0 & 1 & 0 & 0 \\
 0 & -1 & 0 & 0 & 0 \\
 0 & 0 & 0 & 0 & 0 \\
 0 & 0 & 0 & 0 & 0 \\
\end{pmatrix}
\ \ T^2 =\frac{i}{2} \begin{pmatrix}
 0 & 0 & 1 & 0 & 0 \\
 0 & 0 & 0 & 0 & 0 \\
 -1 & 0 & 0 & 0 & 0 \\
 0 & 0 & 0 & 0 & 0 \\
 0 & 0 & 0 & 0 & 0 \\
\end{pmatrix}
\ \ T^3 =\frac{i}{2} \begin{pmatrix}
 0 & 1 & 0 & 0 & 0 \\
 -1 & 0 & 0 & 0 & 0 \\
 0 & 0 & 0 & 0 & 0 \\
 0 & 0 & 0 & 0 & 0 \\
 0 & 0 & 0 & 0 & 0 \\
\end{pmatrix}
\ \ T^4 =\frac{i}{2} \begin{pmatrix}
 0 & 0 & 0 & 0 & 0 \\
 0 & 0 & 0 & 0 & 0 \\
 0 & 0 & 0 & 0 & 0 \\
 0 & 0 & 0 & 0 & -1 \\
 0 & 0 & 0 & 1 & 0 \\
\end{pmatrix} \nonumber \\
\ \ T^5 &=\frac{i}{2} \begin{pmatrix}
 0 & 0 & 0 & 0 & 0 \\
 0 & 0 & 0 & 0 & 1 \\
 0 & 0 & 0 & 0 & 0 \\
 0 & 0 & 0 & 0 & 0 \\
 0 & -1 & 0 & 0 & 0 \\
\end{pmatrix}
\ \ T^6 =\frac{i}{2} \begin{pmatrix}
 0 & 0 & 0 & 0 & 0 \\
 0 & 0 & 0 & 1 & 0 \\
 0 & 0 & 0 & 0 & 0 \\
 0 & -1 & 0 & 0 & 0 \\
 0 & 0 & 0 & 0 & 0 \\
\end{pmatrix}
\ \ T^7 =\frac{i}{2} \begin{pmatrix}
 0 & 0 & 0 & 1 & 0 \\
 0 & 0 & 0 & 0 & 0 \\
 0 & 0 & 0 & 0 & 0 \\
 -1 & 0 & 0 & 0 & 0 \\
 0 & 0 & 0 & 0 & 0 \\
\end{pmatrix}
\ \ T^8 =\frac{i}{2} \begin{pmatrix}
 0 & 0 & 0 & 0 & -1 \\
 0 & 0 & 0 & 0 & 0 \\
 0 & 0 & 0 & 0 & 0 \\
 0 & 0 & 0 & 0 & 0 \\
 1 & 0 & 0 & 0 & 0 \\
\end{pmatrix}\nonumber \\
\ \ T^9 &=\frac{i}{2} \begin{pmatrix}
 0 & 0 & 0 & 0 & 0 \\
 0 & 0 & 0 & 0 & 0 \\
 0 & 0 & 0 & -1 & 0 \\
 0 & 0 & 1 & 0 & 0 \\
 0 & 0 & 0 & 0 & 0 \\
\end{pmatrix}
\ \ T^{10} =\frac{i}{2} \begin{pmatrix}
 0 & 0 & 0 & 0 & 0 \\
 0 & 0 & 0 & 0 & 0 \\
 0 & 0 & 0 & 0 & 1 \\
 0 & 0 & 0 & 0 & 0 \\
 0 & 0 & -1 & 0 & 0 \\
\end{pmatrix}\,.
\end{align}

\end{appendix}